\def\be{\begin{equation}}
\def\te{\end{equation}}
\def\bea{\begin{eqnarray}}

\def\tea{\end{eqnarray}}

\def\a{\alpha}
\def\b{\beta}

\def\d{\delta}

\def\f{\phi}

\def\j{\psi}
    \def\l{\lambda}        \def\o{\omega}
              \def\t{\tau}
       \def\D{\Delta}  
    \def\L{\Lambda}    \def\O{\Omega}

\newskip\humongous \humongous=0pt plus 1000pt minus 1000pt

\newif\ifdtup

\def\ha{{1\over 2}}

\documentstyle[12pt]{article}

\makeatletter                    %allow @ in command names
\@addtoreset{equation}{section}  %reset equation count in each section
\makeatother                     %disallow @ in command names

\begin{document}

\title{Nontrivial Dynamics in the Early Stages of Inflation}

\author{E. Calzetta\thanks{ Email: calzetta@dfuba.edu.ar} and
C. El Hasi\thanks{ Email: elhasi@iafe.edu.ar}\\
{\small IAFE and Department of Physics, University of Buenos Aires,
Argentina}}
%\date{\today}
\maketitle

\begin{abstract}
Inflationary  cosmologies,  regarded  as  dynamical systems,
have rather simple  asymptotic  behavior,  insofar  as  the  cosmic  baldness
principle holds.  Nevertheless,  in  the  early  stages  of  an  inflationary
process, the dynamical behavior may  be very complex.  In this paper, we show
how  even  a  simple  inflationary  scenario,   based  on  Linde's  ``chaotic
inflation''  proposal,  manifests  nontrivial dynamical effects such  as  the
breakup of invariant tori, formation of cantori and Arnol'd's diffusion.  The
relevance of such effects is highlighted by the fact that even the occurrence
or not of inflation in a given Universe is dependent upon them.

\end{abstract}

\newpage

\setcounter{equation}{0}
\section{Introduction}

In this paper, we shall study the behavior of simple  inflationary  models of
the  Universe,  regarded  as  dynamical  systems.  For concreteness, we shall
concentrate on models such as Linde's chaotic inflation scenario\cite{Linde},
where inflation is  powered  by  the vacuum energy of a single slowly rolling
inflaton field.  We  shall also restrict ourselves to the actual inflationary
period, well before ``reheating'' starts.   As it could be expected, we shall
find that, once inflation begins, the Universe quickly falls into a De Sitter
like  expansion,  in  agreement  with  the  ``cosmic    baldness''  principle
\cite{cosmicbaldness}.    Nevertheless,  we  shall  also find that  extremely
complex  behavior  may occur in the very early stages  of  inflation,  before
expansion  becomes exponential.  The list of nontrivial dynamical effects  to
be  found  includes  the  break  up  of  Kolmogorov - Arnol'd -  Moser  (KAM)
tori\cite{KAM},  the  formation  of  cantori  \cite{cantori},  and  Arnol'd's
diffusion\cite{diffusion}.    Indeed,  whether  a  given  Universe  undergoes
Inflation or not depends on these  effects.  Since the complexity of behavior
increases with the number of degrees of  freedom  of  the  dynamical  system,
similar conclusions hold for models based on non  minimal  inflaton  sectors,
and  indeed for any model based on a second  order  phase  transition  and/or
assuming a ``slow roll'' period\cite{extended}.

The  dynamics of cosmological models, allowing for an inflaton field  as  the
only matter present, has been studied by many authors\cite{inflationdynamics}.
In this studies, it  is  customary  to assume homogeneity and isotropy, and a
convenient gauge choice;  the resulting dynamical system has only two degrees
of freedom, e.  g.,the homogeneous  inflaton  field  amplitude $\Phi$ and the
Friedmann - Robertson - Walker (FRW) ``radius  of  the  Universe''  $a$.  The
behavior  of  the  system  is relatively simple, displaying  an  inflationary
attractor in phase space.

Nevertheless,  gravitation  and  a  homogeneous  inflaton cannot be the  only
fields  in  a workable inflationary model.  At some point,  the  inflationary
expansion must ``gracefully exit'' to a radiation dominated FRW epoch.   This
demands  that the Inflaton field must be able to decay into radiation  during
the  ``reheating''  era.    We  do  not have the freedom to assume that  this
radiation field has vanishing  amplitude  prior  to  reheating.   The initial
conditions for the radiation field,  in  the classical regime, are determined
by the earlier quantum and semiclassical eras.

As a matter of fact, simple  quantum  models of the Universe predict that the
radiation field was in its vacuum state at the beginning of the semiclassical
era\cite{HalliHawking}.  However, quanta of the radiation field  are  created
subsequently out of the gravitational field itself, much in  the  same way as
gravitational perturbations are  generated\cite{Grishchuk}.    The  point  is
that,  while  the  radiation    field  would  be  usually  protected  against
cosmological particle creation by conformal  invariance\cite{Parker2}, in our
case, conformal symmetry is broken by  the  non  -  vanishing inflaton vacuum
expectation value, and the radiation coupling to  it.    Moreover,  conformal
symmetry may be broken by other reasons as  well;    for  example,  radiation
quanta  may  be  created  out  of  the  decay  of    primordial  anisotropies
\cite{inflpartcreat}.  We are therefore led to assume a finite  amplitude for
the radiation field at the beginning of the classical regime.

Indeed,  we must adopt for the radiation field the same ``chaotic''  view  of
initial conditions  \cite{Linde} usually applied to the inflaton.  This means
that matter fields  emerge  from  the  semiclassical  era taking uncorrelated
values in different patches  of the Universe;  the dispersion of these values
is such, that they occasionally approach Planckian scales.  As long as we are
interested in the evolution of a small region of the Universe, though, we can
treat the relevant fields as coherent and homogeneous.

In  this  paper,  therefore,  we shall focus  on  the  dynamical  effects  of
including  a  radiation  field  $\j$ on an inflationary  model,  besides  the
homogeneous component of the inflaton field.  ``Radiation'' may  be described
as  a  conformal  scalar field;  however, as we already  remarked,  conformal
symmetry is broken by the non - vanishing expectation value of  the inflaton,
and $\j$ develops a mass through its coupling to it.

In an earlier communication, we have shown that the dynamical system obtained
from coupling a spatially closed,  FRW Universe, to a conformally coupled but
massive scalar field, displays homoclinic chaos\cite{we,kananas}.    While it
would be improper to describe an inflating  cosmology  as ``chaotic'', as the
typical trajectories are unbounded, it should be clear  that  the breaking of
conformal symmetry results in a definite increase in the  complexity  of  the
dynamics.    Concretely,  while  under  conformal  symmetry there is a  sharp
separation  between inflating  and  recollapsing  Universes,  once  conformal
symmetry is broken there  appears  a  full  measure stochastic layer in phase
space, where orbits may either  be trapped forever, or escape and approach De
Sitter expansion.  The reason why recollapse may be avoided, namely, that the
action variable associated to the radiation field  is  no  longer  conserved,
parallels  the  particle  creation  processes  to be found  in  semiclassical
inflationary  models  \cite{Parker2},  and  which  have  a similar, inflation
enhancing, effect\cite{inflpartcreat}.   Indeed, the above action variable is
essentially the particle  number  of the second quantized radiation field, as
defined by the adiabatic particle model\cite{Birrell}.

In this paper we shall consider two different models displaying the influence
of a scalar radiation field on the  expansion  of  the  Universe in the early
stages of Inflation.  In the first model,  the radiation field itself will be
considered homogeneous, along with the metric and the inflaton field.  In the
second  model,  we  shall allow the radiation field to be  constituted  by  a
homogeneous  background  plus  an  inhomogeneous mode.  Of course, the actual
presence of such modes is dictated by the dynamics of particle creation.  The
metric itself will  be  taken always as that of a spatially closed FRW model,
with ``radius of the  Universe'' $a$.  This assumption, which, in the context
of  the  second  model, is  inconsistent  with  Einstein  equations,  may  be
justified physically.  Indeed, our only  concern is to study how the dynamics
of the Universe reacts to an increased  number  of  degrees of freedom.  From
this  point  of view, the consideration of graviton  modes,  along  with  the
inhomogeneous scalar field mode, short of posing the full  field  theoretical
problem, would not bring any new qualitative features\cite{3dcodes}.

It  is  interesting  to  observe  that taking as canonical variable  a  gauge
invariant  quantity built  from  scalar  and  graviton  modes\cite{gaugeinv},
rather than the scalar  field amplitude itself, would lead to similar results
to ours.  Indeed, these  quantities  may  be  described as scalar fields with
time dependent masses \cite{Robert}.

Let us now describe in more  detail  the  models  to  be  studied.  Since the
inflaton field, given the ``slow roll'' assumption,  plays  no dynamical role
in the epoch of interest, we shall trade  it  by  a  fundamental cosmological
constant  $\Lambda$  (the  full  evolution  of a FRW Universe  coupled  to  a
minimal,  massive  scalar  field is analyzed in Ref.  \cite{Raymond}).    The
radiation field $\j$ will be conformally coupled to $a$, but also have a mass
$m$.  The mass $m\sim e\Phi$, where $e$ is the coupling between  inflaton and
radiation, and $\Phi$  the inflaton vacuum expectation value.  The ``charge''
$e$ is bounded by  the  requirement  that  the induced inflaton self coupling
($\sim  e^4$)  should  not be  too  large,  and  $\Phi$  is  bounded  by  the
assumptions of slow roll over, enough  inflation, etc.  This still allows for
values  as  high  as  $e\sim 10^{-3}$, $\Phi\sim  10^2$  (in  natural  units,
$h=c=8\pi G=1$), which gives $m\sim 10^{-1}$\cite{Borner}.

If the radiation field mass is neglected, we  have conformal symmetry and the
system  is  integrable.   The metric and the field  $\j$  (henceforth,  ``the
field'')  are  decoupled, except the metric reacts to the full  field  energy
density  $\rho/a^4$,  $\rho  =(1/2)  (p_{\j}^2+\j^2)\equiv  {\rm  constant}$.
There are two unstable static  solutions,  the  Einstein Universes defined by
$a=\pm  1/2\sqrt{\Lambda}$,  $\rho =1/16\Lambda$.  They  are  joined  by  two
heteroclinic  orbits.   Within the separatrix, defined  by  the  heteroclinic
orbits,  motion  is  quasiperiodic  and  confined  to  invariant   KAM  tori.
Inflating  orbits  are unbounded, and approach asymptotically the stable  and
unstable  manifolds  of the static solutions.  Thus, all inflating  Universes
share the same asymptotic behavior, in agreement with the ``cosmic baldness''
principle.

In  the  conformally symmetric model, therefore, there is a sharp distinction
between  inflationary  and  recollapsing  initial  conditions.    Restricting
ourselves to Universes arising from a Big Bang ($a=0$ at the origin of time),
they inflate if $\rho  >1/16\Lambda$,  and recollapse if $\rho <1/16\Lambda$.
There is no orbit connecting  one  region  to  the  other,  as the separatrix
stands as an unsurmountable barrier.

When the field develops  a  mass,  it  becomes non - linearly coupled to $a$.
This coupling induces internal resonances  between  $a$  and  $\j$, and, as a
consequence,  both  the  separatrix  and underlying  resonant  KAM  tori  are
destroyed.  Instead, there arises a new kind of structure in phase space, the
stochastic  layer\cite{we,kananas}.    The structure  of the layer  is  best
analyzed by means of Poincar\'e sections\cite{Psections}.  It is  found that,
alongside  with  fixed  points  and invariant tori, there is a  new  kind  of
invariant orbit,  the  ``cantori''\cite{cantori}.  Cantori have gaps in them,
which allow for  communication between different parts of the stochastic layer
and the outside.   The  separation between inflating and recollapsing initial
conditions becomes less clearcut:   orbits  starting  from below the original
separatrix may now find their way  through  the  gaps  and  become inflating,
through a process of Hamiltonian diffusion \cite{diffusion}.   The occurrence
of Inflation in a given Universe, therefore, may  depend  upon  a non trivial
dynamical effect such as the break up of KAM tori.

As long as we consider the field $\j$ as homogeneous, there  will be always a
critical   value  of  the  momentum  $p_a$,  such  that  orbits  leaving  the
singularity from  below  this threshold are bound to recollapse.  This is due
to the fact  that  the available phase space, once the Hamiltonian constraint
is  enforced,  is  three  dimensional,  and  thus  it  is  separated  by  two
dimensional KAM tori.  Therefore,  any  unbroken  torus traps the phase space
volume inside it, and makes recollapse  unavoidable.   Moreover, it should be
clear that this critical value will be  close  to  the  separatrix  value  of
$p_a\sim  1/\sqrt{8\Lambda}$  at  $a=0$,  at  least for small  field  masses.
Similarly, while phase volume conservation implies that some orbits  starting
from outside the separatrix must enter the stochastic layer and  be  trapped,
it should be clear that the field mass will not greatly  affect  the behavior
of orbits much above the separatrix.  From these considerations, it could  be
concluded  that  the kind of effects discussed above are associated to rather
exceptional initial conditions.

However, as  we  shall  show  presently, the stochastic layer is localized in
phase space because we have assumed a model with only two degrees of freedom.
Such localization is not  found  in  higher  dimensional  models,  like those
allowing for inhomogeneous fields and  geometry.    The  second  model  to be
considered  in  this  paper,  adding  a  single  inhomogeneous  mode  to  the
homogeneous $\j$ background, constitutes a first step  in  the study of those
more complex models.

The  dynamics  of  non  -  integrable,  higher  dimensional    systems,    is
qualitatively   different  from  that  of  their  two  degrees  of    freedom
counterparts.    In our case, we have a six dimensional  phase  space  $(a,\j
,\j_1,p_a,p_{\j},p_{\j 1})$, where $(\j_1 ,p_{\j 1})$ stand for the amplitude
of the  inhomogeneous  mode  and  its  canonically conjugated momentum.  Even
after enforcement of the Hamiltonian constraint, the available phase space is
five dimensional, and it  is  not  divided  by three dimensional tori.  Thus,
unbroken KAM tori do not  negate  diffussion, and, in principle, a trajectory
beginning  from  $a=0$,  with  arbitrarily  small    $\vert   p_a\vert$,  may
nevertheless find its way beyond the separatrix and inflate.

This  effect  provides  a sort of classical  mechanism  for  ``creation  from
nothing''  \cite{Vilenkin}.   It makes unnecessary to assume  an  unnaturally
high  value  of  the  initial  radiation energy density to  explain  how  the
Universe  could  avoid  recollapsing  much  before inflation had a chance  to
begin.   It  recalls  previous  analysis  of  semiclassical  cosmology, where
particle    creation   has    been    invoked    to    fulfil    a    similar
task\cite{inflpartcreat}.  It also  shows  that  the  non  trivial  dynamical
behavior discussed here is indeed widespread in phase space.

Our goal in this paper  is  to point out the manifestation of the non trivial
dynamical effects described above in numerically  generated  solutions to our
models, both the homogeneous and the inhomogeneous,  higher  dimensional one.
Of    course,   since  Hamiltonian  diffusion  is  such  a    slow    process
\cite{Nekhoroshev},  it  would be extremely hard to follow numerically,  with
proper accuracy, a single diffusing orbit from a neighborhood of  the  origin
in  phase  space until it becomes definitely inflationary.  Instead, we  have
built Poincar\'e  sections  \cite{Psections}  for  both models (in the second
model, where the  Poincar\'e  section  would  be  four  dimensional, we shall
present only three dimensional  projections  of  it),  and  studied the local
Lyapunov exponents  along  selected orbits\cite{Wolf}.

Poincar\'e sections are built by  selecting a given plane on phase space, and
collecting the points where a given  orbit  crosses  this  plane  in a chosen
sense.  The full dynamics induces a ``return map'' on the Poincar\'e section,
both flow and map sharing the same degree  of  complexity from the standpoint
of  Dynamical  Systems  Theory.  Thus, for example, the  orbits  of  the  map
induced by an integrable dynamical system with two dimensional sections  will
be either topological circles or else isolated points;  an orbit  which would
not  fit  in  any  of  these  two types is therefore a strong  indication  of
nonintegrability.  Poincar\'e sections  are  a  specially  valuable  tool  in
investigating generally covariant dynamical systems,  as  they  convey  in  a
simple fashion topological, and therefore gauge invariant, information on the
dynamics\cite{we,kananas}.

In our case, the simpler, purely homogeneous  model  has  a  four dimensional
phase space.  We shall choose the $\j  =0$  plane  to  build  the  Poincar\'e
section;  since the canonical momentum $p_{\j}$ is defined  as  a function of
the  other  variables  by  the Hamiltonian constraint, the section is  two  -
dimensional.

The  second,  inhomogeneous model, would have four dimensional sections.  For
ease of  representation,  it is desirable to reduce the dimensionality of the
section, by imposing  a  second  constraint  ($\j_1=0$,  say) on the selected
points.  By reasons  of  numerical  efficiency, it is necessary to impose the
second constraint only within certain  prearranged  accuracy;    this is like
allowing  the  ``space  of  section''  to    have    a    finite    thickness
\cite{Psections}.    As  a  matter of fact,  in  the  range  covered  by  our
simulations, the shape of the sections is essentially  insensitive to $\j_1$,
so we have simply left it unspecified.  We have found it best to describe the
sections  in  terms  of  coordinates  $(a,  p_a,  j)$.    Here  $j=(1/2\omega
)p_{\j}^2$  ($\omega^2=1+m^2a^2$) is the  adiabatically  invariant  amplitude
associated to the homogeneous mode,  at  the  zero  crossings.    Of  the two
undeterminated  canonical  variables  (the  amplitude  and  momentum  of  the
inhomogeneous mode), one can be deduced from  the Hamiltonian constraint, so
we are losing only one dimension of the  real Poincar\'e section.  As we have
verified, this entails no significant loss of information.

Another technique we have  implemented  to  analyze  the  higher  dimensional
model,  is  to  plot,  in  the  space  of  initial  conditions,  the  region
corresponding  to  orbits  which escape  the  separatrix  in  less  than  500
iterations of the Poincar\'e section return map.    Restricting  ourselves to
Universes  with  vanishing  initial  volume  and  field,  and   imposing  the
Hamiltonian  constraint, the space of initial conditions is two  dimensional;
we  have  chosen  the  initial  values  of $p_a$ and $p_{\psi}$  as  suitable
coordinates.  As befits a chaotic system, the border of the  set  of inflating
trajectories is highly irregular, with thongs of inflating initial conditions
penetrating  the  regular  region,    and  islands  of  regularity  otherwise
surrounded  by unstable solutions.   This  plot  affords  a  glimpse  of  the
structure of the Arnol'd web.

A more common tool for the study  of  non  trivial  dynamical  systems is the
calculation of Lyapunov exponents \cite{Lyapunov}.  These measure the average
unstability  of the orbits of a dynamical system.   If  the  system  executes
bounded motions, then the Lyapunov exponents may be related to the Kolmogorov
-  Sinai  entropy;    in  particular,  positive  Lyapunov's  exponents  are a
sufficient  condition    for    chaos   \cite{Pesin}.    In  our  case,  this
identification cannot be  maintained, since most orbits eventually escape the
separatrix and inflate;  here, the Lyapunov exponent would only relate to the
time constant of the inflationary  exponential  expansion of the Universe, as
the cosmic baldness principle asserts itself.

More relevant to our discussion are therefore the  local  Lyapunov exponents,
which  measure  the average eigenvalues of the linearized evolution  operator
around suitable orbits, over finite parts of it\cite{Wolf}.  Essentially,  we
shall  use  local Lyapunov exponents to obtain a quantitative measure of  the
strong  sensitivity  to  initial  conditions  to  be  found in the Poincar\'e
sections (see  below).    Indeed,  we shall see how orbits with close initial
conditions  are  characterized    by  highly  different  sets  of  exponents,
corroborating the visual information conveyed by the Poincar\'e sections.

To optimize our probing  of  phase  space,  we  shall  take  advantage of the
circumstance that the equations defining  our  dynamical  system  remain well
defined even upon the cosmic singularities $a=0$.  Thus, a given Universe may
be analytically continued beyond the singularity, becoming one in a series of
cosmic episodes.  This series shall extend indefinitely unless, at some link,
the  trajectory avoids recollapse and inflates.  Therefore, an  ``orbit''  of
the recurrence map may actually represent a sequence of many  Universes.  The
important  point  in  the  present  context  is not which of these  Universes
inflate and which do not, but rather the fact that some chains, after several
recollapsing cycles, actually end with an inflationary episode.  This implies
a communication between different  regions  of  phase  space  which  would be
utterly impossible under an integrable dynamics.

Our numerical simulations will offer concrete examples of all  the  kinds of
non - trivial dynamical behavior discussed above.  For the homogeneous mode,
we  shall  see,  through  the  Poincar\'e  sections,  the destruction of the
conformal  KAM  tori,  their  replacement  by  chains  of  islands,  and the
formation of an stochastic sea among the islands.  We shall display concrete
instances of orbits  beginning in the stochastic sea, and eventually leaving
the separatrix.  This  shows  that  the  stochastic sea contains no unbroken
tori, but rather it is composed of cantori layers.  As a check on our codes,
we shall verify the validity of  the cosmic baldness principle for inflating
orbits.

We shall also display two Poincar\'e sections  of  the  inhomogeneous model,
corresponding one to a stable, regular trajectory, and  the  other one to an
unstable  one,  which eventually becomes inflationary.  Comparing these  two
sections, we shall be able to observe how the destruction  of  constants  of
motion  is  reflected  in the topology of the orbit.  The  pervading  strong
sensitivity  to initial conditions is also reflected by the local Lyapunov's
exponents associated  to  each  orbit, which we computed.

{}From the results  of  these  numerical  experiments, we shall conclude that a
realistic inflationary cosmological model, regarded as a dynamical system, is
complex enough to allow for highly nontrivial behavior, and that nonlinearity
is a major force in shaping  the behavior of the model, even where the cosmic
baldness principle holds.  A correct understanding  of the incidence of these
kind  of  phenomena may well illuminate several of  the  standing  issues  in
inflationary cosmology, which are still oftentimes solved by recourse to fine
tuning \cite{finetuning}, as well as help us finding the proper  relationship
between classical, semiclassical, and quantum cosmology.

The paper is organized  as follows.  In next section we introduce our models,
and discuss what may be  expected  on the behavior of their solutions, out of
simple analytical arguments.  In the  following section, we shall present the
results of our numerical simulations.  We  conclude,  in  section  IV, with a
brief discussion of the overall lessons to be learned.

\vfill
\eject
\section{The model}

In this  paper,  we  shall  be  concerned  primarily  with  spatially  closed
Friedmann - Robertson  -  Walker  Universes  coupled  to  scalar  matter  and
radiation.  The metric  takes  the form $ds^2=a^2(\eta)(-d\eta^2+d\sigma^2)$,
where $\eta$ is ``conformal time'',  and  $d\sigma^2$  denotes  an  invariant
metric  on  the  euclidean  three sphere.    The  matter  field,  namely  the
``inflaton'', will be taken as a real  scalar field $\Phi$, minimally coupled
to  curvature,  with effective potential $V(\Phi)$.  ``Radiation''  shall  be
described by a massless real scalar field conformally coupled  to  curvature.
This  means  that  the  Lagrangian  density  shall contain a $(-a^4/12)R(\psi
/a)^2$    term,  $\psi$  being  the  conformally  scaled  scalar  field,  and
$R=(6/a^3)(\ddot a+a)$  the scalar curvature (we shall follow MTW conventions
throughout \cite{MTW} ;   a  dot  denotes  a  conformal time derivative).  The
inflaton and radiation fields are coupled to each other; this
coupling is essential to the ``graceful exit'' from the inflationary phase.

Insofar as FRW symmetry holds, both scalar fields must be homogeneous.  The
basic assumption  of  inflationary  cosmologies  is  that $\Phi$ is an slowly
evolving field whose  effective potential is strictly positive.  This
positive vacuum energy acts as a cosmological constant, powering an explosive
expansion of the Universe (``inflation'').  The nonvanishing background value
of the inflaton also provides a mass to the radiation field, through Higgs'
mechanism.  Until the ``reheating'' phase begins (whereby the inflaton vacuum
energy  is massively transformed into radiation energy and  dissipated
\cite{reheating}),  the
inflaton plays no dynamical role and may be taken  as  constant,  its effects
being incorporated through the values $\Lambda$ of the vacuum energy  density
and $m$ of the radiation field mass.  In this regime,  the  effective degrees
of  freedom  of  the  model are the ``radius of the Universe'' $a$,  and  the
conformally scaled radiation field  $\psi  (\eta)$.  Their conjugated momenta
are just their conformal time derivatives $\dot a=-p_a$, $\dot\psi =p_{\psi}$.
The  evolution of these variables in  conformal  time  is  described  by  the
Hamiltonian

\be
H=({1\over 2})[-(p_a^2+a^2-2\Lambda a^4)+(p_{\psi}^2+\psi^2)+m^2a^2\psi^2],
\label{homham}
\te

\noindent
Supplemented by the Hamiltonian  constraint  $H=0$.    We  shall refer to the
system defined by this Hamiltonian as the ``homogeneous'' model.

The homogeneous radiation field may also be  considered  as the first term in
the expansion of a generic field configuration in  terms of three dimensional
spherical harmonics \cite{3dsphhar}.  The different modes are labelled  by  a
positive integer $n\ge 1$;  $n=1$ corresponds to the homogeneous  mode.  On a
FRW  background,  the  higher modes may be described by their conformal  time
dependent  amplitude    and  its  time  derivative.    Of  course,  when  the
backreaction of these  modes  is  taken  into account, the geometry ceases to
belong to the FRW class.  However, to obtain a glimpse of the effect of these
modes on the cosmic evolution,  we  may disregard the departure of the metric
from its FRW value.  For  simplicity,  moreover,  we  shall assign a non zero
amplitude to only one mode, say $n_1$, with amplitude $\psi_1$ and conjugated
momentum $p_1$. The Hamiltonian now reads

\be
H_1=({1\over 2})[-(p_a^2+a^2-2\Lambda a^4)+(p_{\psi}^2+\psi^2)+
(p_1^2+n_1^2\psi_1^2)+m^2a^2(\psi^2+\psi_1^2)],
\label{inhomham}
\te

\noindent
with,  as  before,  the  constraint  $H_1=0$.    We    shall  refer  to  this
generalization of the model above as the ``inhomogeneous'' one.

If  the  effective mass of the radiation field is  zero,  then  geometry  and
radiation  are  decoupled,  and the resulting dynamics is trivial.   When  a
nonvanishing  mass  develops,  on  the  other  hand,  the  dynamics  of  both
homogeneous and inhomogeneous model becomes exceedingly complex.  Indeed, the
dynamics of the $\Lambda  =0$  homogeneous case has been studied in detail in
ref.  \cite{we}, where it  is demonstrated how homoclinic chaos arises out of
the internal resonances between $a$ and $\psi$.  The $\Lambda\not= 0$ case is
briefly considered in ref \cite{kananas}.  There  it  is  shown  that  in the
massless case there are two unstable static solutions joined by a separatrix;
in  the  presence  of  a nonvanishing mass the separatrix  is  destroyed  and
homoclinic chaos results.  A more detailed analysis based on the structure of
fixed  points and resonances of the homogeneous model confirms these findings
\cite{trio}.

In  view  of  these  results,  it  should  be  clear  that  a    thorough
investigation of these models requires numerical techniques.  However,
perturbative arguments, where the  effective  mass  of the radiation field is
taken as small parameter, shed  some  light on the structure of the solutions
and are helpful to understand the  numerical  results.    Therefore,  in this
section, we shall consider briefly what can  be  said  on these models out of
simple  perturbative  arguments.    Later,  from next section  on,  we  shall
consider several numerical experiments, which will disclose the nature of the
dynamics without any perturbative assumptions.

\subsection{The ``homogeneous'' model}

In  this  subsection,  we  shall  consider  the  dynamics  generated  by  the
two-degrees  of  freedom  Hamiltonian  (\ref{homham}), constrained to the null
energy shell.

In  the  massless case, we may analyze the evolution of  field  and  geometry
independently.  The field is just an harmonic oscillator, with unit frequency
disregarding  its amplitude.  The radius of the Universe, on the other  hand,
may be described as an anharmonic oscillator, whose
 frequency begins as unity, but falls down with amplitude.
It actually  vanishes  as  we  approach  the  separatrix  connecting the two
unstable static solutions.    Beyond the separatrix, motion is unbounded, and
quickly approaches a De Sitter like expansion.

Indeed, the frequency  of  the  field  oscillations  is  weakly dependent on
amplitude as far as  the  mass  remains small.  Therefore, it makes sense to
describe  the field evolution in  terms  of  the  massless  action  -  angle
variables $j$ and $\varphi$, defined through  $\psi  =\sqrt{2j}\sin\varphi$,
$p_{\psi}=\sqrt{2j}\cos\varphi$.      The  Hamiltonian  may  be  split    as
$H=H^0+\delta    H$,   where  the  integrable,  ``unperturbed''  Hamiltonian
$H^0=-E+j$,

\be
E=({1\over 2})[p_a^2+\omega^2a^2-2\Lambda a^4],
\label{E}
\te

\noindent
where $\omega^2=1-m^2j$, while the ``perturbation''

\be
\delta H=({-1\over 2})m^2ja^2\cos 2\varphi
\label{deltaH}
\te

We  shall  retain  the  mass  correction to the small
oscillations frequency, though formally a  small  term,  as  this  shall
improve
remarkably the accuracy of our analysis.

It is clearly seen from (\ref{E})  that  the  properties of the motion depend
strongly on whether $m^2j$ is larger or smaller than unity, in agreement with
previous  results  \cite{we,kananas}.    In the first case,  the  unperturbed
motion  is  unbounded, and all trajectories correspond to a  near  De  Sitter
inflationary  expansion.    Therefore, it is in the opposite case,  $m^2j<1$,
that we may find non trivial dynamical effects.  This is  also  the  relevant
case  to investigate the physical question of whether inflation requires over
planckian radiation energy densities in the early Universe.

Assuming therefore $m^2j<1$, we find
that when $E=E_s=\o^4/16\L$, the ``unperturbed'' motion admits two
static solutions with $a^2=a_s^2=\o^2/4\L$.  These solutions are joined  by a
separatrix. To describe motion below this separatrix, it is
convenient  to  pick  $E$ itself as generalized  momentum.    Its  conjugated
variable is, of course, ``time'', defined as

\be
\tau =\int^a{da'\over\sqrt{2E+2\L a'^4-\o^2a'^2}}
\label{time}
\te

This definition is easily inverted to yield

\be
a=\sqrt{1-\l}a_s{\rm sn} [{\o\t\over\sqrt{1+k^2}}],
\label{ellipticsn}
\te

\noindent
where ${\rm sn}u$ denotes the Jacobi elliptic function of the first kind
\cite{Grad}, $\l =\sqrt{1-(E/E_s)}$, and $k^2=(1-\l )/(1+\l )$
. We can also Fourier expand $a$ as

\be
a=\sqrt{1-\l}a_s\sum_{n=1} q_n\sin (2n-1)\O\t
\label{Fourier}
\te

\noindent
where the  Fourier  coefficients  $q_n$ have standard expressions in terms of
complete elliptic integrals (\cite{Grad}), and the fundamental frequency

\be
\O ={\pi\over 2{\bf K}}{\o\over\sqrt{1+k^2}}
\label{freq}
\te

Here, ${\bf  K}={\bf  K}[k]$ is the complete elliptic
integral of the first kind \cite{Grad}.  We see that $\O\to\o$ when $E\to 0$,
and vanishes when $E$ approaches $E_s$, as we expected.

Let us observe that, because the transformation from $(a, p_a)$ to $(\t ,E)$
depends on $j$, when  the second pair is used, the angle conjugated to $j$ is
no longer $\varphi$ but  $\f =\varphi +\d\f$, where

\be
\d\f ={m^2\over 2}\int^{\t}d\t '~a^2(\t ')
\label{newangle}
\te

Having  solved  the  unperturbed motion, the  analysis  follows  an  standard
pattern  (\cite{kananas}).    The  perturbative  approach  breaks  down,  and
nontrivial  dynamics  appears, when the oscillations in $a$  become  resonant
with those in $\f$.  Since the leading perturbation goes like $\cos 2\f$, the
leading  resonances  will  be  those  where  $N\O  =2$, the unperturbed  $\f$
frequency  being  identically $1$.  This equation has no solution for  $N=1$,
and only the trivial one $E=j=0$ for $N=2$.  Thus the first real breakdown of
perturbation theory  occurs  when  $N=3$.  The corresponding KAM torus in the
unperturbed dynamics is  destroyed by the resonant perturbation, resulting in
new homoclinic points and separatrices.  These separatrices are
further destroyed by other resonant  terms,  leading  to  the formation of an
stochastic layer.  The layers of  different  resonances  tend to overlap with
each  other,  thus  forming an stochastic sea  extending  up  to  the  former
separatrix  at $E_s$.  Within the stochastic sea KAM  tori  are  replaced  by
cantori \cite{cantori},
allowing for such phenomena as diffussion;  thus  an  orbit
with initial conditions below $E_s$ may scape the separatrix and inflate.

The location of the $N=3$ resonance is therefore  an  estimate  of  the lower
limit of the stochastic sea in the full dynamics.  It  is  determined  by the
resonant  condition $\O =2/3$ plus the Hamiltonian constraint $E=j$.  For the
values $m=0.65$, $\L =.125$, which we shall use in our numerical simulations,
the $N=3$ resonance appears at $E=j=0.28$.  For an orbit starting from $a=0$,
this means that  the initial momentum should be $p_a\sim 0.75$.  Moreover, it
being a $N=3$ resonance,  we  expect  that a triangular pattern of islands of
stability will form in the  wake  of  the  destroyed  KAM torus, in $(a,p_a)$
space.

As  we  shall see in next  section,  both  predictions  of  the  perturbative
analysis  are  confirmed  by the numerical results.    For  the  time  being,
however, let us turn to the study of the second, ``inhomogeneous'' model.

\subsection{The inhomogeneous model}

In this subsection, we shall apply essentially the  same  techniques  than in
the previous one, to the study of the ``inhomogeneous'' model \ref{inhomham}.
As before, we extend to all phase space the  low  amplitude  action  -  angle
variables  $(j,\varphi  )$  for  the  ``homogeneous''  component.    For  the
``inhomogeneous'' mode, we write, in the same spirit

\be
\psi_1=\sqrt{{2j_1\over n_1}}\sin\varphi_1
\label{newaction}
\te

\be
p_1=\sqrt{2n_1j_1}\cos\varphi_1
\label{newnewangle}
\te

This allows us to split the Hamiltonian into a perturbation

\be
\delta H_1=-\ha m^2a^2(j\cos 2\varphi +{j_1\over n_1}\cos 2\varphi_1)
\label{newdelta}
\te

\noindent
acting on the unperturbed Hamiltonian $H_1^0=-E+(j+n_1j_1)$, where

\be
E=({1\over 2})[p_a^2+\omega_1^2a^2-2\Lambda a^4]
\label{newE}
\te

\noindent
and $\o_1^2=1-(m^2/n_1)(n_1j+j_1)$.

The integration of the unperturbed motion may be performed exactly as before.
There are two unstable fixed points, at energy  $E=E^1_s=\o_1^4/16\L$, joined
by a separatrix.  Above the separatrix, motion is unbounded.  To describe the
motion  below the separatrix, we write $E=E^1_s (4k^2/(1+k^2)^2)$, $0\le k\le
1$.   The evolution of $a$ may be described as a  superposition  of  harmonic
oscillations, with fundamental frequency $\O_1 =(\pi /2{\bf K})
(\o_1/\sqrt{1+k^2})$.
Here, as before, ${\bf K}={\bf K}[k]$
is the complete elliptic integral of the first kind
\cite{Grad}.

As  in  the  ``homogeneous'' model, it is convenient to choose $E$ itself  as
canonical momentum.  Again this implies introducing a new angle variable $\t$
instead of $a$,  and  shifting  the  angles  $\varphi$ and $\varphi_1$ to new
angles $\f=\varphi +\d\f$ and  $\f_1=\varphi_1+(1/n_1)\d\f$.    The  shift is
still given by the expression  (\ref{newangle}),  where now of course we must
compute the dependence of $a$ on $\t$ using the proper frequency $\o_1$.

Given the structure  of  the  perturbation,  we expect the leading resonances
shall be those where  the  $a$  frequency is a rational multiple of either of
the field frequencies.  Mode  -  mode  coupling, on the other hand, will only
show up at high orders in  perturbation  theory,  and  its  effects  shall  be
correspondingly  weak.

As a  matter  of  fact,  the structure of the resonances, as described by the
``unperturbed''  Hamiltonian  $H_1^0$,    is   rather  trivial.    Since  the
frequencies are independent of  the  field amplitudes, they are just vertical
lines,  accumulating  towards $E=E^1_s$.   Moreover,  since  the  frequencies
belonging to either mode have a fixed, integer ratio, instead of two families
of resonances, there is only one.  To break this degeneracy, we must consider
higher orders in perturbation theory.

To this end, we shall observe that, if we average the perturbation $\d H_1$
over $\t$, holding $\f$, $\f_1$ fixed, we get the non zero value

\be
\langle \d H_1\rangle ={\O_1\over 4\pi}\{ j[\sin 2(\f -\D\f )-\sin 2\f ]
+{j_1\over n_1^2}[\sin 2(\f_1 -{1\over n_1}\D\f )-\sin 2\f_1]\}
\label{avedeltaH}
\te

\noindent
where $\D\f$ is the accumulated phase shift over one period, namely

\be
\D\f ={m^2\o_1\over\L}{[{\bf K}-{\bf E}]\over\sqrt{1+k^2}}
\label{Deltaphi}
\te

\noindent
${\bf E}={\bf E}[k]$ being the complete  elliptic integral of the second kind
\cite{Grad}.  We shall improve our perturbative  scheme  by  introducing this
term into the unperturbed Hamiltonian.  To perform  the  actual  calculation,
however, let us disregard the dependence of $\o_1$ with  respect  to $j$, and
$j_1$;    we  may,  for example, approximate $\o_1^2\sim 1-m^2 E$,  which  is
accurate  when  $j_1\sim  0$.    This  step  may  not  be easily  justifiable
quantitatively, but  it  will  simplify enormously the analysis below without
changing the characther of the solutions.

Treating the new term in the Hamiltonian as small, we seek to elliminate it
by introducing new action variables $x$ and $x_1$, where

\be
j=\{ 1-{\O_1\over 4\pi}[\sin 2(\f -\D\f )-\sin 2\f ]\}x
\label{newx}
\te

\be
j_1=\{ 1-{\O_1\over 4\pi n_1^2}[\sin 2(\f_1 -{1\over n_1}\D\f )-\sin 2\f_1]\}
x_1
\label{newx1}
\te

This transformation is made  canonical by a suitable redefinition of the
angle variables, which we do not need to carry through explicitly.
Writing the Hamiltonian in terms  of  the new action variables, and averaging
over the old angles, we obtain the new, improved, unperturbed Hamiltonian

\be
H_1^1=-E+\a x+n_1\b x_1
\label{newimproved}
\te

\noindent
where

\be
\a=1-2({\O_1\over 4\pi})^2\sin^2\D\f
\label{alpha}
\te

\be
\b=1-2({\O_1\over 4\pi n_1^3})^2\sin^2{\D\f\over n_1}
\label{beta}
\te

It may be observed that  $1-\b\ll 1-\a$ throughout the energy range, so there
is no significative loss in approximating $\b\sim 1$.

To obtain the actual form of  the  resonances,  let  us solve the Hamiltonian
constraint for $x_1$, obtaining

\be
x_1={E-\a x\over n_1}
\label{x1fromHC}
\te

Since $x_1$ must be positive, we must have $x\le E/\a$.

The equations for the two families of resonances now read:

{\sl Case I: a - $\f$ coupling}:

\be
x^I={1\over \a '}\{ 1-{q\a\over\O_1}\}
\label{caseI}
\te

\noindent
where $\a '=d\a /dE$, and $q<1$ is a rational number, and

{\sl Case II: a - $\f_1$ coupling}:

\be
x^{II}={1\over \a '}\{ 1-{q'n_1\over\O_1}\}
\label{caseII}
\te

\noindent
where $q'<1/n_1$  is  also  rational.    We  see that the two
families of resonances have been resolved, although  of course the difference
in slope between the curves of each family are much smaller than the slopes
themselves.   This weak splitting is nevertheless enough to  communicate  the
different  resonances  among  themselves, thus providing the basic pattern of
the Arnol'd web.

Another feature of the resonant lines which can be easily checked against the
numerical results concerns their overall slope.  First, it should be observed
that,  the denominator  in  Eqs.    (\ref{caseI})  and  (\ref{caseII})  being
generally very small over  the  physical range, we only obtain the constraint
$x\le E/\a$ when the numerator  itself  is  close  to  zero.  Moreover, since
$\O_1$ is decreasing over the physical  range, at the zero the numerator will
go from positive to negative. For the values $m=.65$ and $\L
=.125$ to be used in next section,  the  denominator is positive over
the range $.237\le E\le .355$. Since $x$ must be positive, we may conclude
that the resonant lines have negative slope in this range of energies, reaching
the zeroes of the numerator from below.   In  terms  of  the initial value of
the momenta, assuming initially $a,\f ,\f_1=0$, this interval corrresponds to
$.688\le  p_{ai}\le  .842$. Of course, a negative $dx/dE$ also implies $dp_{\j
i}/dp_{ai}\le  0$,  in  agreement  with  the  numerical  results (see below).
Outside this range, the resonances should display positive slopes; however,
this  behavior  has  not  been observed, as motion remains regular below  the
lower limit, and it is too unstable above the upper one.

We shall stop our analysis at this point, proceeding to  the presentation
of the numerical results.

\section{Numerical Results}

We now proceed to present the results of the numerical solution of the models
described in Section II.  In these numerical simulations, we have reverted to
the    original    canonical   variables  $(a,p_a  ,\psi  ,p_{\psi},  \psi_1,
p_1)$, whose  equations  of  evolution  follow  from  the Hamiltonians
(\ref{homham}) and (\ref{inhomham}).  The simplicity of these equations makes
this approach more appealing  than  other,  more sophisticated, alternatives.
We have also appealed to  the  fiction of continuing each solution beyond the
cosmic  singularities $a=0$.  Therefore, the  numerical  trayectories  to  be
presented correspond actually to strings of different  Universes,  each being
the analytic extension of its parent. This  mathematical trick,
which is made possible by the fact that the equations of motion, as written in
the  conformal  time  frame,  remain  analytical at these singularities, will
allow for a more efficient exploration of the dynamics, without affecting the
physics.

To solve  the  models,  we have used a Runge Kutta \cite{NumRecipes} $5^{th}$
order routine.   This method works rather well in our examples, mainly due to
the particular choice of  time  variable.   We have chosen not to enforce the
Hamiltonian Constraint directly, but rather,  as  a  check  on  the numerical
code, surveyed it in all runs,  making sure that the value of the Hamiltonian
never exceeded a given threshold of $10^{-10}$.

Let us now discuss our results.   As in the previous section, we shall begin
with the simplest, ``homogeneous'' model.

\subsection{Homogeneous Model}

This  section is devoted to study  a  homogeneous  inflationary  cosmological
model consistent in a FRW background, an  inflaton  field  and  a conformally
coupled  scalar  radiation  field.  Because we are  interested  in  dynamical
effects during the very beginning of the inflationary stage,  we will replace
the  inflaton  field by a cosmological constant $\Lambda$, also allowing  the
radiation field to develop a mass $m$ through Higgs' mechanism.   Here and in
next section, we assume $\L =.125$ and $m=.65$, both in natural units.  These
values are  suitable  both  on  theoretical  grounds  and  for  the practical
implementation of the  model;    moreover,  it  can  be seen that the general
features of the solutions are independent of this particular choice.

Using standard  dynamical  systems theory techniques the 4-dimensional phase
space of this  model,  can  be  reduced to a more visualizable 2-dimensional
Poincar\'e surface of section.    This  is  built  by intersecting a plane (
$\psi = 0$, say), with  the  dynamical flow in the phase space.  In this way
all the dynamics can be analized  on,  for  example,  the $a$ - $p_a$ plane.
These variables represent the ``slow'' variables in our problem.

If the conformal symetry  were  not  broken  by  the  presence  of mass, the
problem would be integrable and there would be a separatrix sharply dividing
the Poincar\'e section in five regions.    One  region,  around  the origin,
where the motion is bounded, is filled  with  invariant KAM tori.  The other
ones  correspond  to  unbounded  trajectories that, wheresoever they  begin,
approach a De Sitter solution.  The motion would  be  regular  in  all  five
regions.

Due to the mass term the separatrix is replaced by  a  stochastic  layer.  In
this  layer  the  tori, corresponding to the inner region, are broken,  being
replaced by new structures, cantori and islands of stability.  The former are
partially broken  tori  with  gaps  in  them;    some trajectories can escape
through the gaps,  while  others remain confined within the stochastic layer.
Trapped trajectories also form around the elliptic points left in the wake of
resonant tori, and make up the islands of stability.

The five regions on the plane of  section  are  displayed  in Fig.  (1).  The
solid  lines  represent  the  separatrices connecting the massless  unstable
static solutions

\be
p_a = \pm {1 - 4 \Lambda a^2 \over \sqrt{8 \Lambda} }
\label{separatrix}
\te

\noindent
as well as the other branches of the stable and  unstable  manifolds emerging
from  the  massless  fixed  points,  which  represent  inflationary De Sitter
solutions.  We have also plotted the sections belonging to a few trajectories
in the massive case, which manage to escape from the separatrices.  It can be
easily appreciated that all  the orbits approach very rapidly the massless De
Sitter solution.  This is,  of  course,  just what would be expected from the
``cosmic baldness'' principle.  We can  also  observe  that  the separatrices
shrink in the presence of mass, as  may  be expected from the analysis in the
previous section.

The fact that some orbits manage to get through a region  formerly  occupied
by  KAM  tori is a clear indication of the formation of cantori.    However,
since  for  small  masses  the  gaps  in the cantori are rather narrow, most
orbits escape  only  after  a  long period of bouncing within the stochastic
layer, while others  remain  bouncing  forever.  Due to Liouville's theorem,
some other orbits must  get in from outside the separatrix, being trapped in
the stochastic sea.

The  region within the separatrices is shown in greater detail  in  Fig  (2).
This  picture  corresponds  to almost a hundred different initial conditions,
all of  them  with  $a_i  = \psi_i = 0$, but differing in $p_{\psi i}$, which
runs up  to  0.812.    As  mandated  by  the  Hamiltonian constraint,
$p_{ai}=-p_{\psi i}$.  We  may  appreciate the inner unbroken KAM tori, where
the points seem to lie  on smooth curves.  We note that they are not arranged
in sequential order, so each torus  corresponds  to  several turns around the
origin, or, in cosmological language, to several  cosmic cicles.  After those
there are several `orbits' corresponding to broken tori, which after several
revolutions escape and inflate. Outermost,
a new set of stable trajectories makes up a triangular pattern of islands.
This
chain  of secondary islands sorrounds the central one in a rather  symetrical
way.  The slight asymmetry of  the  two  lower islands, due to their relative
position with respect to the most likely escape route,
can  be  attributed to the particular choice of initial conditions.
It is easily  seen  that  as  the mass is increased the last unbroken KAM torus
shrinks, but, as long as the field amplitude is different from
zero, it never reaches the origin of  phase  space.  The approximate location
of  the  islands of stability, as well as  the  triangular  pattern,  are  in
excellent agreement with the analysis in the previous section.

In Fig.  (3), we exhibit a blow up  of the upper island.  At the bottom there
are several unbroken tori, then a chaotic or irregular region, and the stable
island surrounding an elliptic point on the former KAM torus.   As it is well
known,  in  non integrable systems there exists an infinity hierarchy of such
structures, in such a way that around a secondary island there is a  pattern
of  third  order  islands,  around each of these islands exist a fourth order
chain of  islands  and  so  on.  The radial width of the corresponding island
decreases with the order of the chain.  This kind of behaviour can be clearly
appreciated in the picture,  where up to third order islands may be seen.  If
we were sufficiently clever as to get initial conditions close enough to some
resonance, it would be possible to  go  on  forever, bringing to ligth deeper
and deeper levels in phase space, until  eventually  the  resolution  of  the
computer is exhausted.

We find that not all the orbits in  the  chaotic  region  escape to infinity.
Indeed,  some  of them remain in the stochastic layer  for  more  than  $300$
iterations of the map, a rather stable behaviour.  Also, there is no clearcut
divide between stable and unstable trajectories;  as we increase the  initial
momenta, there are unstable orbits following stable ones and viceversa.

As another way to stress the variety of behaviour, we have computed the local
Lyapunov characteristic numbers  or  Lyapunov  exponents, after the algorithm
proposed  by Benettin and  others  \cite{Wolf},  for  different  trajectories
corresponding to an unbroken tori,  broken  tori leading to irregular motion,
and stable orbits within the upper  secondary island.  Fig.  (4) is a plot of
the  largest Lyapunov Number corresponding to an  unestable  trajectory  with
initial conditions:  $\psi_i = a_i = 0, p_{\psi i} =-p_{ai}= 0.736$ .  After an
initial
transitory stage, the Lyapunov coefficient becomes positive for several  tens
of recurrencies on the Poincar\'e section.  It never stabilizes,  and towards
the end it sharply grows,  due,  most  likely,  to the unbounded character of
this orbit .  On the other hand, for initial conditions within  the  large
central island or some of the secondary ones, the Lyapunov
characteristic numbers tend  slowly to
zero after the  initial transitory stage, as expected for regular trajectories.

To  summarize,  this simple model displays extremely complex behavior.    The
fact that the main features and location of the caothic region answers to the
analysis  in  Section  II indicates that these are legitimate effects, rather
than  numerical  constructs.    The  dynamical effects to be seen include the
destruction of  KAM  tori,  and  their  replacement  by cantori and stability
islands.

With regards to  the  cosmological  relevance  of these findings, it could be
objected that they are restricted to a rather special region of phase space.
To counter this argument  we  need  to consider the second, ``inhomogeneous''
model, to which we presently shift our attention.

\subsection{Inhomogeneous Model}

In this section we shall  add a second, inhomogeneous  mode, to the radiation
field.  In this way  we  will  be  able to analyze the reaction of the system
under an increase in the number  of  degrees of freedom.  Indeed, phase space
has now 6 dimensions $(a,p_a ,\psi ,p_{\psi},  \psi_1,  p_1)$;   if we
take  a section, due to the fixed value  of  the  Hamiltonian,  we  obtain  a
4-dimensional  Poincar\'e space of section, which is much more  difficult  to
visualize than the 2-dimensional surfaces of section of the previous case.

In  the  case  of  an  integrable  system  there  are,  in  addition  to  the
Hamiltonian, two other  integrals  of  motion.    Therefore  the  orbits  are
confined to three dimensional  tori,  and  their  sections  can  be contained
within 2-dimensional subsets of the  space  of  section.    If  the system is
perturbed, but we are in a region where the KAM theorem applies, then we also
expect 2-dimensional sections.  At the other  extreme, for an ergodic system,
points should fill the 4-dimensional space of section.    Thus, the dimension
of the manifold occupied by a sequence of points  belonging to the same orbit
can range from 2 to 4.

In  our  specific model, if the radiation field were masless the system would
be  integrable.  For example, we could  choose  as  constants  of  motion  in
involution  the  Hamiltonian itself and the field action  variables  $j$  and
$j_1$, introduced in the previous Section.  When the mass is turned on, these
variables are no longer conserved;  however, away from resonances,  we expect
there will be two other constants of motion, analytic in $m^2$,  and reducing
to  $j$,  $j_1$  in  the  massless  limit.    Therefore,  in  the absense  of
resonances, we expect that the orbits will leave a two dimensional imprint
on a given Poincar\'e space of section, which we shall choose, by analogy with
the previos section, as the $\j =0$ plane.

To  display this section, however, it  is  convenient  to  project  it  onto
ordinary  three  dimensional space by constraining also  one  of  the  other
``fast'' variables (\cite{henon}), say $\j_1$.  In fact,  we  have  found in
our numerical trials that the resulting projection is largely insensitive to
the chosen interval in $\j_1$;  thus, for simplicity, we  shall leave $\j_1$
unspecified,  projecting  the actual section onto $(a, p_a , p_{\j})$ space.
Of course, in the massless case, the projection lies entirely on a  $j=~{\rm
constant}$ surface.  In the massive case,  but in the absense of resonances,
there will be an invariant surface close to  this plane, which shall contain
the orbit.  Therefore, the Poincar\'e section will bend,  but  it  will
preserve its shape.  On the other hand, upon a  resonance,  the section will
break apart altoghether.

To  better  appreciate this effect,  it  is  convenient  to  elliminate  the
adiabatic distortion of the section produced  by the evolution of the radius
of the Universe $a$.  To this end, we shall display the numerically obtained
sections in $(a, p_a, J)$ space, where

\be
J={\{ p_{\j}^2+(1+m^2a^2)\j^2\}\over 2\sqrt{1+m^2a^2}}
\label{adiabaction}
\te

\noindent
is  an  adiabatic invariant.  Even with this  choice  of  variables,  motion
becomes highly irregular as soon as we approach the  separatrices.  For this
reason,  we  shall  confine  ourselves  to the region in phase  space  where
nonintegrable  behavior  is  first  apparent.  Also, we shall concentrate on
orbits where  $\j_1$ and $p_1$ are (initially) small, since our main concern
is to study  how  the  presence  of  the  ``inhomogeneous'' mode affects the
dynamics of $a$ and $\j$.

Fig.5 shows a very stable, non resonant trajectory of our inhomogenous model,
it corresponds to  more  than  a  thousand  recurrencies  of the orbit on the
Poincar\'e section.  The initial conditions for it are:  $\psi_i = \psi_{1 i}
= a_i = 0, p_{\psi  i} = 0.7293, p_{ai} = - 0.7295 $ and $p_{1i}\sim 0.0171$,
as given by the Hamiltonian constraint.    All  the  trajectory  seems  to be
confined just to a line;   although  a few
points are seen away from the main line of  section,  they  correspond to the
late  behavior  of  the  simulation,  where  the  loss  of stability  may  be
attributed to  the  accummulation of numerical errors through the simulation.
It is also  remarkable  that  the  section  does  not  display  gaps;  we are
therefore seen an unbroken KAM torus, rather than a cantori.

Now, Fig.6 shows the corresponding picture, for an orbit characterized by the
neighboring initial conditions ($\psi_i = \psi_{1i} = a_i  =  0, p_{\psi i} =
0.7293, p_{ai} = -0.7294$ and $p_{1i}\sim 0.0121$).  This  orbit is much less
stable  than  the  previous  one,  becoming inflationary after a few  hundred
iterations of the map.  Besides, it is hard to believe that it could fit into
a line;  it seems rather to belong to a surface or  even  a  volumen.    This
orbit therefore  corresponds  to  a broken torus.  It becomes inflationary by
diffussing around the  unbroken  tori  surrounding  it, rather than by moving
accross gaps in them

To  corroborate  the  results above, Fig.7 displays a comparision between the
local Lyapunov  exponents  associated  to  each  trajectory.    The upper one
corresponds to the  second  orbit  and its value is around twice the value of
the first one.  The first orbit is clearly more stable despite the fact it is
closer to the separatrices.   It therefore suggest that the unstable orbit is
able to somehow find its way around the stable ones, becoming inflationary in
spite of the survival of unbroken tori  in the outer layers of the stochastic
sea.  Indeed, we expect unstable orbits to  ``diffuse''  along  the  resonant
lines of the Arnol'd web.  This provides a  classical  mechanism  to  achieve
inflation starting from fairly generic initial conditions.

Of  course,  to  obtain  a  direct map of the web  is  exceedingly  difficult
(\cite{diffusion}),  since  a  typical  unstable orbit only stays within it a
relatively small fraccion of the time.  Indeed, typical evolution is made  of
short  bursts  of fast diffusion along the web, followed by long intervals of
seemingly stable motion.    There are in the literature rigorous upper bounds
for the actual diffussion  rates  (\cite{Nekhoroshev});    they are generally
exponentially small.

We may obtain an idea of the structure of the web, however, by studying which
initial conditions lead  to stable (resp.  unstable) motion.  By constraining
ourselves to initial conditions  with  $a_i=\j_i  =\j_{1i} =0$, and enforcing
the Hamiltonian constraint, we may  reduce  these  initial  conditions to the
$(p_{ai}, p_{\j i})$ plane.  Initial conditions close to the web will diffuse
more easily and become inflationary faster.   Thus, by mapping the regions of
instability, we obtain a rough idea of the outline of the web.

In Fig.   8  we  display  the  sector $0.7288\le p_{ai}\le 0.7308$, $0.725\le
p_{\j i}\le 0.73$ in  the  plane  of  initial conditions.  We have considered
initial conditions evenly distributed in  this  sector, marking with an ``*''
those which inflated in less than  500  iterations of the map (``500 ticks of
the $\j$ clock''), and with a ``-''  those  which  did  not.  Of course, only
initial conditions with $p_{ai}\ge p_{\j 1}$ are physical;    we have removed
the lower left and upper right corners, where no structure can be seen.

The map shows clearly the intrincate structure of the border of chaos region.
We can see lines of instability running from upper  left to lower right (just
as expected from the analysis in Section II).  Of course,  what we are seeing
is not just the resonant lines, but the stochastic layers around them.  As we
progress  towards the upper right, these layers merge, and the stochastic sea
is formed.  Towards the upper left of the figure, where  the  orbits in Figs.
5  and 6 belong, stable orbits predominate, but it is easy to  discern  paths
connecting unstable initial  conditions  to the stochastic sea, moving around
the stable islands. These are the presumed pathways to inflation.

Fig. 9 is a blow up of the central region of Fig. 8. Here,
$0.7294\le  p_{ai}\le  0.7298$,  $0.727\le   p_{\j  i}\le  0.728$.    We  see
essentially no less structure than  in  the  larger  plot,  which recalls the
fractal nature of the actual  web.    As in the earlier case, the survival of
certain  islands of stability has no  confining  effect  on  deeper  unstable
orbits.

We thus see that the inhomogeneous model is not only more ``chaotic'' than
its  homogeneous counterpart, which after all was to be  expected,  but  also
that  the ``fine tuning'' objection to the cosmological relevance of  chaotic
behavior is  essentially  groundless.    In  higher  dimensional  models, the
survival of KAM  tori  does not confine the deeper layers of phase space, and
so essentially all of  phase space is connected to the chaotic region.  Quite
to  the contrary, it is the  existence  of  caothic  diffusion  which  makes
it
possible to dispose of the assumption of an unnaturally high radiation energy
density to avoid recollapse before inflation.   Of  course,  the  addition of
semiclassical  effects  such  as  particle  creation  only  reinforces   this
conclusion \cite{inflpartcreat}.

\section{Final Remarks}

The  main  goal  of  this  paper has been to  show  that  even  the  simplest
cosmological models are liable to display highly nontrivial dynamics.
This  is  so  even in models, such as the ones we have considered
here, where the  cosmic  baldness  principle  severely  limits the asymptotic
behavior.  The consequences  of  this  behavior  for  the  evolution  of  the
Universe may well be drastic;  in the models we have studied, the realization
or not of Inflation could hinge  upon the possibility of chaotic diffusion in
phase space.

A second lesson to be learned is the clear difference of behavior between the
model  with  two degrees of freedom, and that  with  three.    It  should  be
stressed  that from a purely dynamical point of view,  the  addition  of  the
inhomogeneous mode is a rather mild change, since it does not couple directly
to  the  homogeneous  mode,  and its larger proper frequency shields it  from
resonances  with the ``radius of the Universe'' (cfr.  section II).  However,
the purely  topological  change  brought  by the opening of new dimensions in
phase space is  enough  to  produce a marked increase in complexity, of which
the ``deconfinement'' of unstable  orbits  near  the origin of phase space is
the most conspicuous.

This  indicates  that  Galerkin    (\cite{Lichtenberg})   or  Minisuperspace
\cite{Misner}  type  approximations  should  be   used  with  extreme  care,
specially when addressing the dynamics of  the  early Universe, when motions
were fast and parametric couplings strong.   This word of caution, which has
recently  came  to  be  accepted  by  researchers  of    the    quantum  era
\cite{Kuchar}, is not less true in classical cosmology.

\section{Acknowledgments}

This work has been partially supported by Fundaci\'on
Antorchas, CONICET, and UBA (Argentina).

We are grateful to David Hobill and H\'ector Vucetich for discussing with us
an earlier version of this work.

%\newpage

\section{Figure Captions}
\medskip
\noindent
{\bf Figure 1} Poincar\'e section on  the  $\psi  =  0$  plane.   Only the no
boundary trajectories are shown.  The solid  lines  belong  to  the conformal
symetric case ($m = 0$).  Notice the  running away trajectories approache the
separatrix.

\smallskip
\noindent
{\bf Figure 2}  Same  as  figure  1,  including  the  region of quasiperiodic
motion.  There are  more  than a hundred trajectories that start at $\psi_i =
a_i = 0$ , with  $p_{\psi  i}$  running  up  to  $\sim  0.812$.    It  can be
appreciated the three secondary islands.

\smallskip
\noindent
{\bf Figure 3} Blow up of the upper island of figure 2.  Around this island a
similar pattern of third order islands is clearly seen.

\smallskip
\noindent
{\bf  Figure  4}  Maximal  Lyapunov exponent for an unstable trajectory (with
initial conditions:    $\psi_i  =  a_i  = 0$ , and $p_{\psi i} = 0.736$).  It
never stabilizes and  grows up suddenly when the system escapes away from the
separatrix.

\smallskip
\noindent
{\bf Figure 5} Poincar\'e space of section for a stable trajectory (more than
a thousand iterations on the  section),  for  the  inhomogeneous  model.  The
initial conditions are:  $\psi_i = \psi_{1 i} = a_i = 0, p_{\psi i} = 0.7293,
p_{ai} = - 0.7295 $ and $p_{1i}\sim  0.0171$,  as  given  by  the Hamiltonian
constraint.

\smallskip
\noindent
{\bf Figure 6} Same as figure 5 for an unstable trajectory.  In this case the
initial conditions are:  $\psi_i = \psi_{1i} = a_i  = 0, p_{\psi i} = 0.7293,
p_{ai}  =  -0.7294$  and  $p_{1i}\sim  0.0121$.  Despite the lower  value  of
$p_{ai}$ this is a less regular orbit than the former.

\smallskip
\noindent
{\bf Figure 7} Comparison  between the local Lyapunov exponents associated to
trajectories shown in figures 5  and  6.    The  upper one corresponds to the
second orbit and its value is  around  twice the value of the first one.  The
first orbit is clearly more stable despite  the fact that it is closer to the
separatrix.

\smallskip
\noindent
{\bf Figure 8} The sector $0.7288\le p_{ai}\le 0.7308$, $0.725\le p_{\j i}\le
0.73$ in the plane of  initial  conditions.    Marked with an ``*'' are those
trajectories which inflated in less than  500  iterations  of  the map (``500
ticks of the $\j$ clock''), and with a ``-'' those which did not.  Of course,
only initial conditions with $p_{ai}\ge p_{\j i}$ are physical.

\smallskip
\noindent
{\bf Figure 9} A blow up of the central region of figure 8.  Here, $0.7294\le
p_{ai}\le  0.7298$,  $0.727\le    p_{\j  i}\le  0.728$.    The  structure  is
essentially the same as  in the larger plot, which recalls the fractal nature
of the actual web.   Also  in  this  case, the survival of certain islands of
stability has no confining effect on deeper unstable orbits.

\end{document}